\documentclass[lettersize,journal]{IEEEtran}
\usepackage{amsmath,amsfonts}
\usepackage{algorithmic}
\usepackage{array}
\usepackage[caption=false,font=normalsize,labelfont=sf,textfont=sf]{subfig}
\usepackage{textcomp}
\usepackage{stfloats}
\usepackage{url}
\usepackage{verbatim}
\usepackage{graphicx}
\def\BibTeX{{\rm B\kern-.05em{\sc i\kern-.025em b}\kern-.08em
    T\kern-.1667em\lower.7ex\hbox{E}\kern-.125emX}}
\usepackage{balance}
\usepackage{booktabs}
\usepackage{cite}
\begin{document}
\title{Physics-Guided Diffusion Transformer with Spherical Harmonic Posterior Sampling for High-Fidelity Angular Super-Resolution in Diffusion MRI}

\author{
Mu Nan, Taohui Xiao, Ruoyou Wu, Shoujun Yu, Ye Li, Hairong Zheng,  \\  and Shanshan Wang, \IEEEmembership{Member, IEEE}
\thanks{Manuscript received **; accepted **. This work was supported in part by the National Natural Science Foundation of China under Grant 62222118 and Grant U22A2040,  in part by Shenzhen Medical Research Fund under Grant B2402047, in part by Key Laboratory for Magnetic Resonance and Multimodality Imaging of Guangdong Province under
Grant 2023B1212060052, and in part by the Youth Innovation Promotion Association CAS. (Corresponding author: Shanshan Wang,e-mail: sophiasswang@ hotmail.com)}
\thanks{Mu Nan, Taohui Xiao, Ruoyou Wu, Shoujun Yu, Ye Li, Hairong Zheng, and Shanshan Wang are with the Paul C. Lauterbur Research Center for Biomedical Imaging, Shenzhen Institutes of Advanced Technology, Chinese Academy of Sciences, Shenzhen 518055, China}
}

\maketitle
\begin{abstract}
Diffusion MRI (dMRI) angular super-resolution (ASR) aims to reconstruct high-angular-resolution (HAR) signals from limited low-angular-resolution (LAR) data without prolonging scan time. However, existing methods are limited in recovering fine-grained angular details or preserving high fidelity due to inadequate modeling of q-space geometry and insufficient incorporation of physical constraints. In this paper, we introduce a Physics-Guided Diffusion Transformer (PGDiT) designed to explore physical priors throughout both training and inference stages. During training, a Q-space Geometry-Aware Module (QGAM) with b-vector modulation and random angular masking facilitates direction-aware representation learning, enabling the network to generate directionally consistent reconstructions with fine angular details from sparse and noisy data. In inference, a two-stage Spherical Harmonics-Guided Posterior Sampling (SHPS) enforces alignment with the acquired data, followed by heat-diffusion-based SH regularization to ensure physically plausible reconstructions. This coarse-to-fine refinement strategy mitigates oversmoothing and artifacts commonly observed in purely data-driven or generative models. Extensive experiments on general ASR tasks and two downstream applications, Diffusion Tensor Imaging (DTI) and Neurite Orientation Dispersion and Density Imaging (NODDI), demonstrate that PGDiT outperforms existing deep learning models in detail recovery and data fidelity. Our approach presents a novel generative ASR framework that offers high-fidelity HAR dMRI reconstructions, with potential applications in neuroscience and clinical research.
\end{abstract}

\section{Introduction}
Diffusion magnetic resonance imaging (dMRI) has emerged as a powerful non-invasive tool to probe tissue microstructure and connectivity by measuring the displacement of water molecules under diffusion-sensitized gradients\cite{jian2007unified}. In particular, high angular resolution (HAR) diffusion imaging extends conventional diffusion tensor imaging by sampling along a large number of gradient directions, which enables the reconstruction of complex fiber orientation distributions and more accurate estimates of microstructural metrics \cite{du2011diffeomorphic,hedouin2020interpolation}. These advances have proven critical for a wide range of downstream tasks, including diffusion tensor imaging (DTI), Neurite Orientation Dispersion and Density Imaging (NODDI) for clinical diagnostics and neuroscience research\cite{le2001diffusion,zhang2012noddi}. However, accurate estimation of the above applications typically relies on acquiring HAR dMRI data across multiple b-values and dozens of diffusion directions, which translates into long scan times that are often prohibitive in clinical settings, especially for patient populations with limited tolerance or in multi-site studies where throughput and cost are major concerns \cite{assaf2005composite,jeurissen2013investigating}. Traditional q-space interpolation approaches solve this by fitting continuous basis functions to the measured q-space samples, most prominently using spherical harmonics (SH). SH interpolation models the diffusion weighted (DW) signal as a series of SH coefficients. To further reduce interpolation artifacts, they often use physically plausible smoothing like Gaussian kernels or heat kernels that is equivalent to applying isotropic heat diffusion on the sphere\cite{seo2010heat, neuman2013tikhonov}. Since the dMRI signal evolves according to the heat diffusion equation on the sphere, these techniques yield smooth signals without spikes\cite{chung2015unified,neuman2013tikhonov}. While useful to some extent, the spherical interpolation frameworks still degrade substantially when only a limited number of diffusion directions are available \cite{caiazzo2016q}.

To address these challenges, deep learning–based (DL) methods have been explored to reconstruct HAR diffusion signals directly from sparsely sampled low angular resolution (LAR) DWIs, a task often referred to as angular super‐resolution (ASR) in dMRI. \cite{zeng2022fod,lyon2022angular}. Among them, both physical-informed convolutional neural networks (CNN) and generative models achieved great success via integrating dMRI-specific prior knowledge into the deep learning models' inherent characteristics\cite{lyon2022angular, zhao2024super,ye2019deep,zhang2024phy}. Previous state-of-the-art (SOTA) CNN–based approaches often treat the series of diffusion measurements across q‐space gradient directions as a sequential signal, leveraging their capacity to model complex spatial correlations \cite{lyon2022angular,lyon2023spatio, ye2020improved}. However, the intrinsic smoothing operations of CNNs and the common practice of training on pre‐denoised data often limit their ability to restore subtle microstructural features under low signal‐to‐noise‐ratio (SNR) conditions. Recently, generative-frameworks offer a powerful probabilistic modeling for noise‐like textures and capturing heterogeneity over diffusion directions at fine scales\cite{ren2021q,chen2024qid,zhang2024phy}, yet they frequently lack sufficiently stringent explicit data‐consistency constraints, which can lead to hallucinated artifacts that undermine clinical reliability. Furthermore, large‐scale evaluations have demonstrated that diffusion signal representations learned on one acquisition protocol do not reliably transfer across scanners or b‐value schemes\cite{de2021generalizability}. Data‐driven DL methods may generalize poorly to higher b‐value or multi‐shell data when the SNR is low\cite{sabidussi2023dtirim}. Overall, the detailed angular reconstruction, rigorous adherence to the measured DWI signals and robust performance across high b-value acquisitions remain challenging.

Compared with the mentioned approaches, the proposed method focuses more on introducing the dMRI-specific physical guidance actively into both training and inference, addressing limitations in fine‑grained detail recovery and data fidelity. In this paper, we introduce Physics-Guided Diffusion Transformer (PGDiT) tailored for general ASR in dMRI. Our key contributions are:
\begin{itemize}
    \item A unifying physics guidance paradigm is introduced for the diffusion transformer for ASR in dMRI. Compared with existing methods that often rely only on learned priors or smoothing without fidelity constraints, the proposed method explicitly integrates physical guidance during both large-scale pretraining and diffusion reverse sampling, enabling q-space representation learning to improve the recovery of fine-grained details and the consistency with condition partial DWI samples.
    \item A q-space geometry-aware module (QGAM) based on the feature-wise modulation mechanism is proposed specifically for the transformer backbone. By embedding diffusion gradient directions, QGAM enables the model to explicitly leverage physical orientation information during training, improving its ability to reconstruct subtle angular features.
    \item An spherical harmonics-guided diffusion posterior sampling (SHPS) is employed during inference phase to ensure sampling fidelity, with a coarse SH estimate for initial guidance and a heat-diffusion based SH coefficient regularization as a smooth search. This coarse-to-fine optimization ensures hard data-consistency and physically grounded smoothness, enhancing robustness across varied b-values regimes.
    
\end{itemize}

\section{Related Works}
There has been an increasing amount of research employing DL-based methods for ASR in dMRI. Considering the technical aims, we will review the current research status from two aspects of task-specific and general ASR methods accordingly.

\subsection{Task-driven Deep Learning-Based ASR Methods}
Within this data-driven ASR paradigm, one branch of work is explicitly tailored to downstream applications where the network is trained end-to-end to predict specific diffusion‐derived parameter maps or metrics from undersampled acquisitions \cite{golkov2016q,ye2017tissue,ye2019deep,ye2020improved,qin2021super,qin2021multimodal}. It has been demonstrated that embedding richer data priors into network architectures can markedly improve downstream microstructure estimation from severely under‐sampled diffusion weighted imagings (DWI). For example, Golkov et al. first showed that a simple multilayer perceptron (MLP) could predict key diffusion parameters from highly undersampled data, recovering scalar maps such as mean diffusivity and fractional anisotropy with reasonable accuracy\cite{golkov2016q}. Building on this, Ye et al. introduced a two-stage architecture for NODDI model fitting, leveraging sparsity priors in the diffusion signal to refine parameter estimates in a cascaded manner \cite{ye2017tissue}. Subsequently, they further incorporated spatial context into the q-space deep learning (q-DL) framework—first by embedding neighboring voxel information to stabilize microstructure estimates\cite{ye2019deep}, and later by integrating a hierarchical feature extractor to capture complex tissue heterogeneity\cite{ye2020improved}. Qin et al. advanced this line of research by proposing the super-resolved q‐DL (SR-q-DL) method, which enhances the fidelity of microstructural parameter maps through a residual learning scheme, and further introduced a probabilistic SR-q-DL variant to quantify uncertainty in the network’s outputs\cite{qin2021super,qin2021multimodal}. These task‐driven frameworks collectively underscore the potential of DL methods to deliver clinically practical, high‐quality diffusion metrics from vastly reduced acquisition schemes. Despite these successes, task-driven ASR methods share a fundamental limitation: because each network is optimized for a specific downstream metric or model (e.g., DTI tensors, NODDI parameters), it must be redesigned and retrained whenever a new diffusion model or clinical application is targeted. 
\begin{figure*}[!ht]
\centerline{\includegraphics[width=\linewidth, height=0.4\textheight, keepaspectratio]{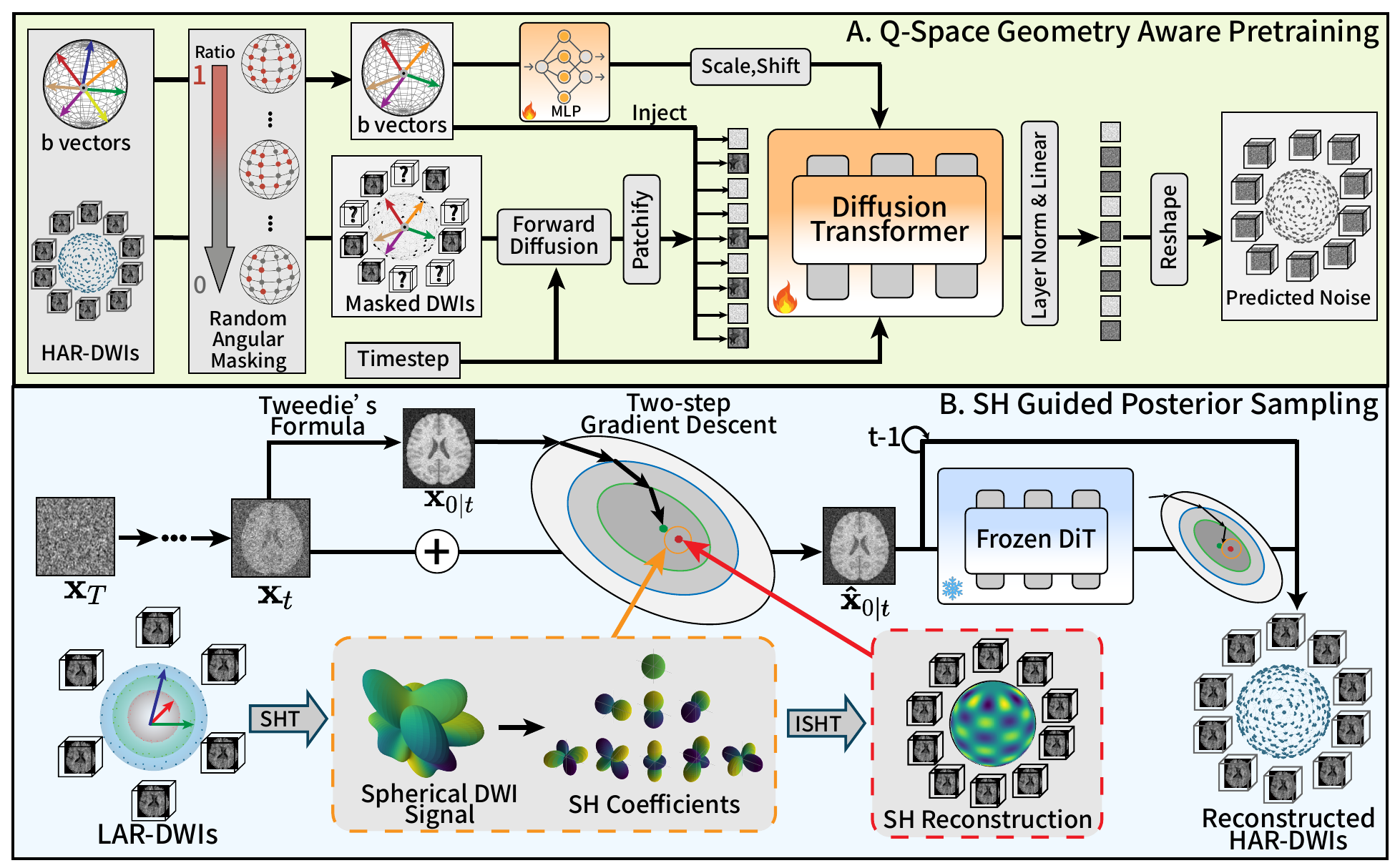}}
\caption{
Overview of the proposed PGDiT framework.
(A) Q-Space Geometry-Aware Pretraining: HAR DWI volumes and their corresponding b-vectors are processed using random angular masking to simulate LAR inputs. The masked DWI tokens, combined with positional encodings and b-vector–based scale-shift modulation (generated via an MLP encoder), are injected into a diffusion transformer.
(B) SH-Guided Posterior Sampling: During inference, the pretrained transformer is frozen and used to perform reverse diffusion. At each sampling step, first a coarse reconstruction is obtained via SH fitting (red dot), and second, the heat diffusion based smoothing is imposed to encourage convergence toward physically plausible smooth solutions (yellow circle), towards the target HAR DWI signals (green dot).
}
\label{Archi}
\end{figure*}
\subsection{General Deep Learning ASR Methods}
An alternative to task-driven approaches is to perform super‐resolution directly on the LAR DWIs themselves, enabling broad compatibility with a wider range of downstream analysis\cite{lyon2022angular,lyon2023spatio,ren2021q,chen2024qid,zhang2024phy}.  Lyon et al. proposed a 3D recurrent convolutional neural network (3DRCNN) conditioned on target gradient vectors, using ConvLSTM blocks for patch‐wise regression in q‐space\cite{lyon2022angular}. Building on this, the same group developed a parametric continuous convolution network (PCConv) that embeds Fourier feature mappings and domain‐specific context into a continuous kernel framework achieving competitive accuracy\cite{lyon2023spatio}. These two CNN–based frameworks first apply  denoising or smoothing to their training data in order to suppress noise, but this comes at the cost of attenuating the high‐frequency angular information that is crucial for resolving complex fiber configurations\cite{lin2022magnitude}.
Recently, generative models have been developed to model noise‐like textures and recover fine-grained details\cite{karimi2024diffusion}.  Ren et al. introduced generative adversarial networks (GAN)‐based frameworks conditioned on b‐values and b‐vectors, augmented with T1‐ and T2‐weighted images to guide generation of raw DW signals \cite{ren2021q}. More recently, Chen et al. employed an image‐conditioned diffusion denoising probabilistic model (DDPM)-based generative model with a U‐Net backbone and cross‐attention to preserve positional cues when upsampling in q‐space \cite{chen2024qid}. Despite the improved recovered details via DDPM's strong capability of recovered details, it does not leverage physics‐informed priors for signal smoothness and exert no explicit data consistency term. Similarly, Phy‐Diff has incorporated dMRI physical principles into a DDPM framework to guide synthesis \cite{zhang2024phy}. It still falls short in enforcing strict data consistency, which can introduce artifacts especially under low‐SNR conditions. 

\section{Methods}
This section details the architecture, training and inference strategy of PGDiT for ASR in dMRI. As illustrated in Fig.\ref{Archi}, our model comprises two complementary stages. During the large-scale pretraining, inspired by feature-wise modulation and masked modeling\cite{perez2018film, gao2023masked}, we employed the QGAM and angular masking to enable the model to learn robust q-space representations. During diffusion reverse sampling, inspired by SH physics, and posterior sampling\cite{chung2022diffusion}, we designed SH-guided two-stage posterior sampling to enforce explicit measurement consistency.
\subsection{Q-space Geometry Aware Module}\label{sec:qgam}
To incorporate explicit diffusion gradient direction information, we introduce a directional conditioning mechanism using per-direction b-vector embeddings. Each DWI acquisition is paired with a set of b-vectors $\mathbf{b} \in \mathbb{R}^{ N \times 3}$,  where $N$ denotes the number of directions and each vector encodes the diffusion gradient direction in Cartesian coordinates. These b-vectors are processed through a lightweight MLP module termed the Q-space Geometry-Aware Module (QGAM), based on feature-wise linear modulation. Specifically, each b-vector $\mathbf{b}_{n} \in \mathbb{R}^3$ is mapped via an MLP encoder to a pair of per-direction FiLM modulation parameters: 
\begin{equation}
    (\gamma_{i,n}, \beta_{i,n}) = \mathrm{QGAM}\left( \mathbf{b}_{i,n} \right), \quad \gamma_{i,n}, \beta_{i,n} \in \mathbb{R}^D
\end{equation}
where $i = 1, \dots, B$ denotes the batch size, $n = 1, \dots, N$, and $D$ the hidden feature dimension. These coefficients are used to modulate the scale and shift terms of the adaptive LayerNorm-Zero. We Then split these outputs into six components for attention and MLP modulation:
\begin{equation}
    (\hat{\gamma}_{\text{attn}}, \hat{\beta}_{\text{attn}}, g_{\text{attn}}, \hat{\gamma}_{\text{mlp}}, \hat{\beta}_{\text{mlp}}, g_{\text{mlp}}) = \mathrm{Split}_6(\hat{\gamma}_{i}, \hat{\beta}_{i})
\end{equation}
and inject the b-vector–specific modulation direction-wise by broadcasting. These terms are applied to modulate the normalized input patches $\mathbf{h} \in \mathbb{R}^{B \times N \times P \times D}$ at each direction using affine transformations:
\begin{equation}
\begin{aligned}
    \mathrm{Mod}_{\text{attn}}(\mathbf{h}^{(i,n)}) & = \gamma_{\text{attn}}^{(i,n)} \odot \mathbf{h}^{(i,n)} + \beta_{\text{attn}}^{(i,n)}\\
    \mathrm{Mod}_{\text{mlp}}(\mathbf{h}^{(i,n)}) & = \gamma_{\text{mlp}}^{(i,n)} \odot \mathbf{h}^{(i,n)} + \beta_{\text{mlp}}^{(i,n)}
\end{aligned}
\end{equation}
with final outputs gated by learnable per-layer scalars $g_{\text{attn}}, g_{\text{mlp}} \in \mathbb{R}^D$ , and added to the residual connection. This module introduces angular direction awareness into both attention and feed-forward layers, aligning the internal feature representation with the spherical geometry of q-space and improving q-space representation learning ability.
\subsection{Self-Supervised Angular Masking Pretraining}\label{sec:pretraining}
To enable the q-space representational learning, we adopt a random angular masking strategy to efficiently pretrain the model. During training, we randomly omit subsets of diffusion directions, masking a portion of angular tokens and feeding only the remaining visible embeddings—augmented with global and relative angular positional encodings—to the network. Given the inherently long sequences in q-space (e.g. $N \sim 90$), we adapt the transformer by concatenating observed and noisy tokens at the token level, enabling joint modeling of both present and missing inputs.
Let $\boldsymbol{x}_0 \in \mathbb{R}^{H \times W \times C}$ denote the fully sampled DWI images, We then corrupt the masked image ${\boldsymbol{x}}_0$ with a variance schedule $\{\beta_t\}_{t=1}^T$ analogous to DDPMs:
\begin{equation}
    \boldsymbol{x}_t \sim q\bigl(\boldsymbol{x}_t \mid \boldsymbol{x}_0, t\bigr)
    = \mathcal{N}\bigl(\boldsymbol{x}_t;\,\sqrt{\bar{\alpha}_t}\,\boldsymbol{x}_0,\,(1-\bar{\alpha}_t)\mathbf{I}\bigr),
\end{equation}
where $\bar\alpha_t = \prod_{s=1}^t (1 - \beta_s)$. Let $M(k) \in \{0,1\}^{H\times W}$ be a mask operator broadcasting across directions with masking ratio $k$. The masked input is defined as:
\begin{equation}
    \boldsymbol{x}_{\mathrm{inp}}
    = M(k)\!\odot\!\boldsymbol{x}_0 
      \;+\;\bigl( 1-M(k)\bigr)\!\odot\!\boldsymbol{x}_t,
\end{equation}
where $\odot$ denotes element-wise multiplication, and $x_t$ denotes the noisy samples. The denoising network $f_\theta$ is a Transformer that learns to predict the noise $\epsilon$ conditioned the corrupted input, diffusion gradient direction $b$, the mask operator $M(k)$, and the timestep $t$:
\begin{equation}
\epsilon_\theta
= f_{\theta}\bigl(\boldsymbol{x}_t,\,\boldsymbol{x}_0,\,t,\,M(k),b\bigr)\
\end{equation}
In implementation, $\boldsymbol{x}_t$ and $M(k)$ are each partitioned into non-overlapping patches, linearly projected to token embeddings, concatenated with timestep embeddings, and processed through standard Transformer layers similar to the masked diffusion transformer pretraining regime.
We train the model to minimize the expected denoising error across randomly sampled timesteps and masking configurations:
\begin{equation}
\mathcal{L}_{\mathrm{mask}}
= \mathbb{E}\Bigl\lVert\epsilon \;-\;\epsilon_{\theta}\bigl(\sqrt{\bar\alpha_t}\,\tilde{\boldsymbol{x}}_0 + \sqrt{1-\bar\alpha_t}\,\epsilon,\;t,\;M(k),\;b\bigr) \Bigr\rVert^2
\end{equation}
During pretraining, we gradually increase the mask ratio $k$ from a maximum $k_{\min}=0.5$ up to $k_{\max}=0.94$, ensuring that the model gradually encounters more masked contexts as training progresses, to accelerate convergence and improve generalization on non‐convex objectives. 
\subsection{Spherical Harmonics–guided Posterior Sampling}\label{sec:SHPS}
To recover further ensure sampling's fidelity and physical smoothness with the sparse acquisitions of LAR DWIs, we employed a two-stage spherical harmonics–guided posterior sampling. 
\subsubsection{SH-Constrained Observation Consistency}\label{sec:1ststage}
In the first stage of SHPS, we aim to enforce consistency between the model’s denoised prediction and the actual acquired sparse dMRI measurements, while ensuring that the reconstructed signals lie on the manifold of physically plausible diffusion signals.
Given the noisy state \(x_t\), we first compute the denoised estimate \(x_{0|t}\) via Tweedie’s formula:
\begin{equation}
x_{0|t}(x_t)
= \frac{1}{\sqrt{\bar{\alpha}_t}}
\Bigl(x_t \;-\;\frac{\beta_t}{\sqrt{1 - \bar{\alpha}_t}}\;\epsilon_\theta(x_t, t)\Bigr),
\label{eq:tweedie}
\end{equation}
We fuse the observed sparse measurements with the model's estimate, we construct a hybrid signal:
\begin{equation}
\hat{x}_{\mathrm{0}} \;=\; M(k)\odot x_{0} \;+\;(1 - M(k))\odot x_{0|t},
\end{equation}
We then fit SH coefficients to this fused estimated signal:
\begin{equation}
\mathbf{\hat{c}}_{0}
=\arg\min_{\mathbf{c}}\;\bigl\|Y_{\mathrm{obs}}\mathbf{c} \;-\;\hat{x}_{\mathrm{0}}\bigr\|_2^2,
\end{equation}
where \(Y_{\mathrm{obs}}\) is the SH basis matrix evaluated at the acquired b-vectors. This allows us to reconstruct a full HAR DWI signal by projecting $\mathbf{\hat{c}}_{0}$ back to all spherical directions via $Y_{\mathrm{full}}\,\hat{\mathbf{c}}$
To ensure that the denoised estimate adheres to this physically constrained representation, we define the Observation Consistency Loss at step $t$ as
\begin{equation}
\mathcal{L}_{\mathrm{OC}}(x_t) = \| Y_{\mathrm{full}} \mathbf{\hat{c}}_{0} - x_{0|t} \|_2^2
\end{equation}
,which penalizes any deviation of the model’s output from the SH-reconstructed signal, effectively constraining the model to remain within the subspace defined by smooth, physically meaningful diffusion profiles.

\subsubsection{SH-Domain Smoothness Regularization}
In the second stage SHPS, we enforce physical smoothness on the model's denoised estimates by incorporating a heat diffusion-based regularization prior in SH domain. Specifically, low-order SH coefficients capture coarse diffusion structure, while higher-order coefficients encode finer microstructural details but are highly sensitive to measurement noise. To suppress such noise and encourage smoothness, we impose a Laplace–Beltrami regularization on the SH coefficients, which acts as Tikhonov low‐pass filter on the $\mathbb{S}^2$ \cite{neuman2013tikhonov,neuman2012laplace,reuter2009laplace}.
The Laplace-Beltrami operator $\Delta_{\mathbb{S}^2}$ on the unit sphere $\mathbb{S}^2$ acts on spherical harmonics $Y_{l,m}(\theta,\phi)$ as eigenfunctions with eigenvalues $-l(l+1)$. Therefore, applying it to a spherical signal $f(\theta,\phi)$ from yields:
\begin{equation}
\Delta_{\mathbb{S}^2}f = -\sum_{l,m} l(l+1)c_{l,m} Y_{l,m}.
\end{equation}
The corresponding regularization term, becomes:
\begin{equation}
\mathcal{R}(f) = \frac{1}{2}\int_{\mathbb{S}^2}\lVert\nabla_{\mathbb{S}^2}f\rVert^2 d\Omega = \frac{1}{2} \sum_{l,m} l(l+1) c_{l,m}^2.
\end{equation}
which penalizes high‐frequency components proportionally. To integrate this smoothness prior into diffusion inference, we introduce a Signal Smoothness Conservation (SCC) step. Let $C_{\mathrm{pred}}$ be the SH coefficients fit to the fused model estimate and observation $\hat{x}_{\mathrm{0}}$, and let $C_{\mathrm{obs}}$ be the original observed LAR DWI's coefficients. We define the SCC loss as
\begin{equation}
\mathcal{L}_{\mathrm{SCC}} = \Bigl\lVert C_{\mathrm{obs}} - C_{\mathrm{pred}}\Bigr\rVert^2.
\end{equation}
Although this is an $L_2$ loss in SH space, it implicitly enforces heat diffusion based regularization because discrepancies in higher-order coefficients contribute more to the angular mismatch. This second-stage update ensures that the model’s prediction not only aligns with acquired measurements but also conforms to the physically smooth structure of the angular diffusion signal.

In conclusion, during each reverse diffusion sampling iteration, we compute the gradients of $\mathcal{L}_{\mathrm{OC}}$ and $\mathcal{L}_{\mathrm{SCC}}$ and combine them as:
\begin{equation}
\nabla_t = \lambda_{\mathrm{OC}}\,\nabla_{x_t}\!\mathcal{L}_{\mathrm{OC}}
+ \lambda_{\mathrm{SCC}}\,\nabla_{x_t}\!\mathcal{L}_{\mathrm{SCC}}
\end{equation}
, where $\lambda_{\text{OC}}$ and $\lambda_{\text{SCC}}$ are tunable hyperparameters that balance fidelity and smoothness, selected via grid search during inference-time respaced sampling. The final sampling update is performed by subtracting the combined gradient from the current state:
\begin{equation}
x_{t-1} = x_t
\;-\;\lambda_{\mathrm{OC}}\,\nabla_{x_t}\!\mathcal{L}_{\mathrm{OC}}
\;-\;\lambda_{\mathrm{SCC}}\,\nabla_{x_t}\!\mathcal{L}_{\mathrm{SCC}}
\end{equation}

\section{Experiments}

\begin{figure*}[htbp]
\centering
\includegraphics[width=\linewidth, height=0.3\textheight, keepaspectratio]{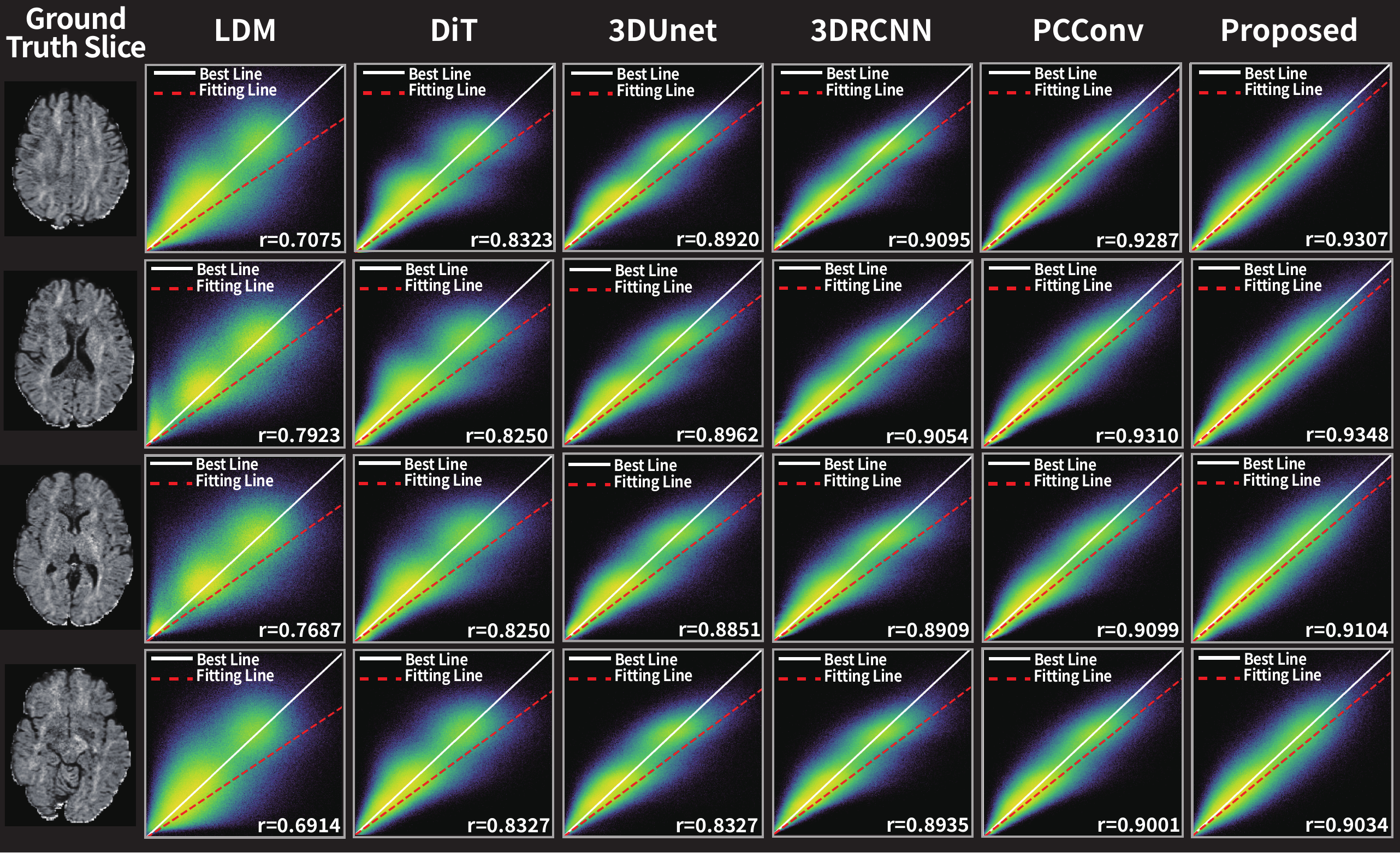}
\caption{
Scatter plot comparison of reconstructed vs. ground-truth DWIs across models. Each scatter plot represents one model’s reconstruction across all predicted diffusion directions from a randomly selected subject's slice. The x-axis corresponds to ground-truth DWIs, while the y-axis shows the model's reconstructions. The white solid line indicates perfect reconstruction, and the red dashed line is the best linear fit. Pearson’s correlation coefficient \(r\) is reported at the bottom of each plot. 
}
\label{fig:scatter}
\end{figure*}
\subsection{Datasets and Preprocessing}
The Human Connectome Project (HCP) Young Adult 3T dataset was used to validate the effectiveness of our method. Data were acquired using a standard 32-channel Siemens receive head coil\cite{van2013wu}. Each scan included three diffusion shells $b=1000,2000,3000s/mm2$, each sampled along 90 diffusion gradient directions, along with 18 non-diffusion-weighted b0 volumes. The resulting 4D diffusion-weighted images had dimensions of 145×174×145×288 with 1.25 mm isotropic resolution. All DWI images underwent official preprocessing through the HCP pipeline. Subsequently, we normalized the DWI volumes by dividing them by the mean of the b0 images. Each subject’s data were separated into the three shells plus the 18 b0 volumes, with each shell comprising 90 directions. For this study, we used data from 130 subjects,where we randomly selected 100 subjects for training, 20 for validation and 10 for testing.

We further included DTI and NODDI models to assess clinically meaningful analysis and evaluation on downstream tasks. To generate reference DTI maps, including mean diffusivity (MD), axial diffusivity (AD), and fractional anisotropy (FA), diffusion tensor fitting was performed using only the $b=1000s/mm2$ shell and b0 images via the DIPY toolkit \cite{garyfallidis2014dipy}. This produced ground truth maps for MD, AD, and FA. In addition, tissue microstructure metrics from the NODDI model were considered, specifically the intra-cellular volume fraction ($V_{ic}$), cerebrospinal fluid (CSF) volume fraction ($V_{iso}$), and orientation dispersion (OD). These gold-standard tissue microstructure maps were computed using the AMICO\cite{daducci2015accelerated} with the full set of 270 diffusion gradients.

To evaluate ASR reconstruction performance at multiple super-resolution scales $r$. Here, the ASR scale $r$ is denoted as  $r = \frac{q_{\mathrm{target}}}{q_{\mathrm{in}}}$, where $q_{\mathrm{in}}$ is the number of input directions and $q_{\mathrm{out}}$ is the number of ground truth HAR DWIs. For each shell in the test set, gradient directions were downsampled using MRITool \cite{cheng2017single}, producing inputs with $q_{\mathrm{in}}$ = 15, 10, and 6 directions, and corresponding ASR scales of $r = 6,9$ and $15$.
\subsection{Implementation Details}
We evaluated our method against several SOTA DL–based general ASR approaches. These include: 3DUNet \cite{suzuki2024high} (implemented via the MONAI framework \cite{cardoso2022monai}), 3DRCNN \cite{lyon2022angular}, PCConv \cite{lyon2023spatio}, latent diffusion model (LDM) \cite{pinaya2022brain}, and the original DiT) \cite{peebles2023scalable}. To note, the original DiT model was trained as a baseline without our proposed QGAM and SHPS. All models were trained on two NVIDIA Tesla A100 GPUs. Unless otherwise specified, we used each method’s recommended hyperparameters for a fair comparison. Training was performed for 200k iterations, using the AdamW optimizer with a learning rate of \(1\times10^{-5}\) and a decay of \(1\times10^{-2}\). All weights were initialized from a standard normal distribution. All models except 3DRCNN were implemented in PyTorch \cite{paszke2019pytorch}; 3DRCNN was implemented in TensorFlow \cite{abadi2016large}. Unless otherwise noted, models were trained to minimize the mean squared error (MSE) loss. Performance was quantitatively assessed using peak signal-to-noise ratio (PSNR) and structural similarity index measure (SSIM). The PGDiT code will be made publicly available upon acceptance.
\begin{table*}[!ht]
\centering
\caption{QUANTITATIVE COMPARISONS OF HAR DIFFUSION IMAGES FROM DIFFERENT METHODS ARE SHOWN ACROSS THREE SHELLS And THREE ASR SCALES. BOLD NUMBERS INDICATE THE BEST RESULTS}
\label{tab:dwi}
\resizebox{0.95\textwidth}{!}{
\begin{tabular}{ccllllll}
\toprule[1pt]
\noalign{\vskip 0.5ex}
\textbf{b value}&  & \multicolumn{2}{c}{$\mathbf{1000s/mm^{2}}$} & \multicolumn{2}{c}{$\mathbf{2000s/mm^{2}}$} & \multicolumn{2}{c}{$\mathbf{3000s/mm^{2}}$} \\[0.5ex] \hline
\noalign{\vskip 0.5ex}
\textbf{ASR Scale} & \textbf{Methods} & \multicolumn{1}{c}{\textbf{SSIM}} & \multicolumn{1}{c}{\textbf{PSNR}} & \multicolumn{1}{c}{\textbf{SSIM}} & \multicolumn{1}{c}{\textbf{PSNR}} & \multicolumn{1}{c}{\textbf{SSIM}} & \multicolumn{1}{c}{\textbf{PSNR}} \\[0.5ex] \midrule[1pt]
\noalign{\vskip 0.5ex}
 & 3DUnet & $0.9356\pm0.0079$ & $27.4947\pm0.4541$ & $0.9155\pm0.0097$ & $27.2262\pm0.5459$ & $0.9061\pm0.0103$ & $26.9868\pm0.6304$ \\
 & 3DRCNN & $0.9542\pm0.0052$ & $28.5647\pm0.3629$ & $0.9422\pm0.0070$ & $28.0038\pm0.5657$ & $0.9250\pm0.0096$ & $27.4381\pm0.6733$ \\
$r=15$ & PCConv & $0.9449\pm0.0055$ & $28.4422\pm0.4445$ & $0.9317\pm0.0077$ & $27.8036\pm0.5009$ & $0.9174\pm0.0091$ & $27.2115\pm0.6132$ \\
 & LDM & $0.8724\pm0.0220$ & $23.7471\pm0.7324$ & $0.8442\pm0.0108$ & $24.6054\pm0.4338$ & $0.8377\pm0.0115$ & $24.2197\pm0.3183$ \\
 & DiT & $0.9195\pm0.0060$ & $26.8805\pm0.3753$ & $0.8989\pm0.0101$ & $26.6302\pm0.4451$ & $0.8948\pm0.0099$ & $26.6136\pm0.5544$ \\
 & Proposed & $\mathbf{0.9569\pm0.0054}$ & $\mathbf{28.5968\pm0.4655}$ & $\mathbf{0.9467\pm0.0065}$ & $\mathbf{28.0721\pm0.5739}$ & $\mathbf{0.9332\pm0.0081}$ & $\mathbf{27.4751\pm0.7392}$ \\[0.5ex] \midrule[1pt]
\noalign{\vskip 0.5ex}
 & 3DUnet & $0.9445\pm0.0069$ & $28.0880\pm0.4702$ & $0.9242\pm0.0088$ & $27.5542\pm0.5749$ & $0.9130\pm0.0094$ & $27.1031\pm0.6369$ \\
 & 3DRCNN & $0.9618\pm0.0024$ & $29.0161\pm0.1394$ & $0.9480\pm0.0067$ & $28.2159\pm0.6035$ & $0.9321\pm0.0088$ & $27.5043\pm0.7226$ \\
$r=9$ & PCConv & $0.9621\pm0.0049$ & $29.0465\pm0.2112$ & $0.9420\pm0.0073$ & $28.2810\pm0.5615$ & $0.9282\pm0.0090$ & $27.5173\pm0.6629$ \\
 & LDM & $0.8923\pm0.0127$ & $24.8061\pm0.4851$ & $0.8625\pm0.0134$ & $25.2333\pm0.5238$ & $0.8556\pm0.0144$ & $24.9904\pm0.4188$ \\
 & DiT & $0.9263\pm0.0056$ & $27.3481\pm0.3810$ & $0.9044\pm0.0094$ & $26.8650\pm0.4569$ & $0.9009\pm0.0092$ & $26.7294\pm0.5635$ \\
 & Proposed & $\mathbf{0.9624\pm0.0049}$ & $\mathbf{29.0646\pm0.4832}$ & $\mathbf{0.9501\pm0.0071}$ & $\mathbf{28.3290\pm0.5462}$ & $\mathbf{0.9369\pm0.0080}$ & $\mathbf{27.5745\pm0.7126}$ \\[0.5ex] \midrule[1pt]
 \noalign{\vskip 0.5ex}
 & 3DUnet & $0.9539\pm0.0059$ & $28.5517\pm0.4874$ & $0.9349\pm0.0077$ & $27.7331\pm0.5782$ & $0.9245\pm0.0084$ & $27.3610\pm0.6624$ \\
 & 3DRCNN & $0.9606\pm0.0050$ & $29.4521\pm0.4430$ & $0.9492\pm0.0066$ & $28.4257\pm0.6196$ & $0.9342\pm0.0086$ & $27.6130\pm0.7671$ \\
$r=6$ & PCConv & $0.9641\pm0.0037$ & $\mathbf{29.7514\pm0.4170}$ & $0.9446\pm0.0064$ & $28.0404\pm0.3872$ & $0.9347\pm0.0084$ & $27.5442\pm0.7212$ \\
 & LDM & $0.9039\pm0.0106$ & $25.4187\pm0.4099$ & $0.8772\pm0.0119$ & $25.5297\pm0.5301$ & $0.8708\pm0.0127$ & $25.6804\pm0.4278$\\
 & DiT & $0.9311\pm0.0055$ & $27.6806\pm0.3787$ & $0.9117\pm0.0087$ & $27.0691\pm0.4654$ & $0.9071\pm0.0086$ & $26.9198\pm0.5789$ \\
 & Proposed & $\mathbf{0.9655\pm0.0046}$ & $29.4183\pm0.4996$ & $\mathbf{0.9520\pm0.0063}$ & $\mathbf{28.4335\pm0.6105}$ & $\mathbf{0.9386\pm0.0074}$ & $\mathbf{27.6289\pm0.6945}$ \\[0.5ex] \bottomrule[1pt]
\end{tabular}
}
\end{table*}
\subsection{Comparisons of the SOTA Models}
\begin{figure}[htbp]
\centering
\includegraphics[width=1\linewidth, height=0.4\textheight, keepaspectratio]{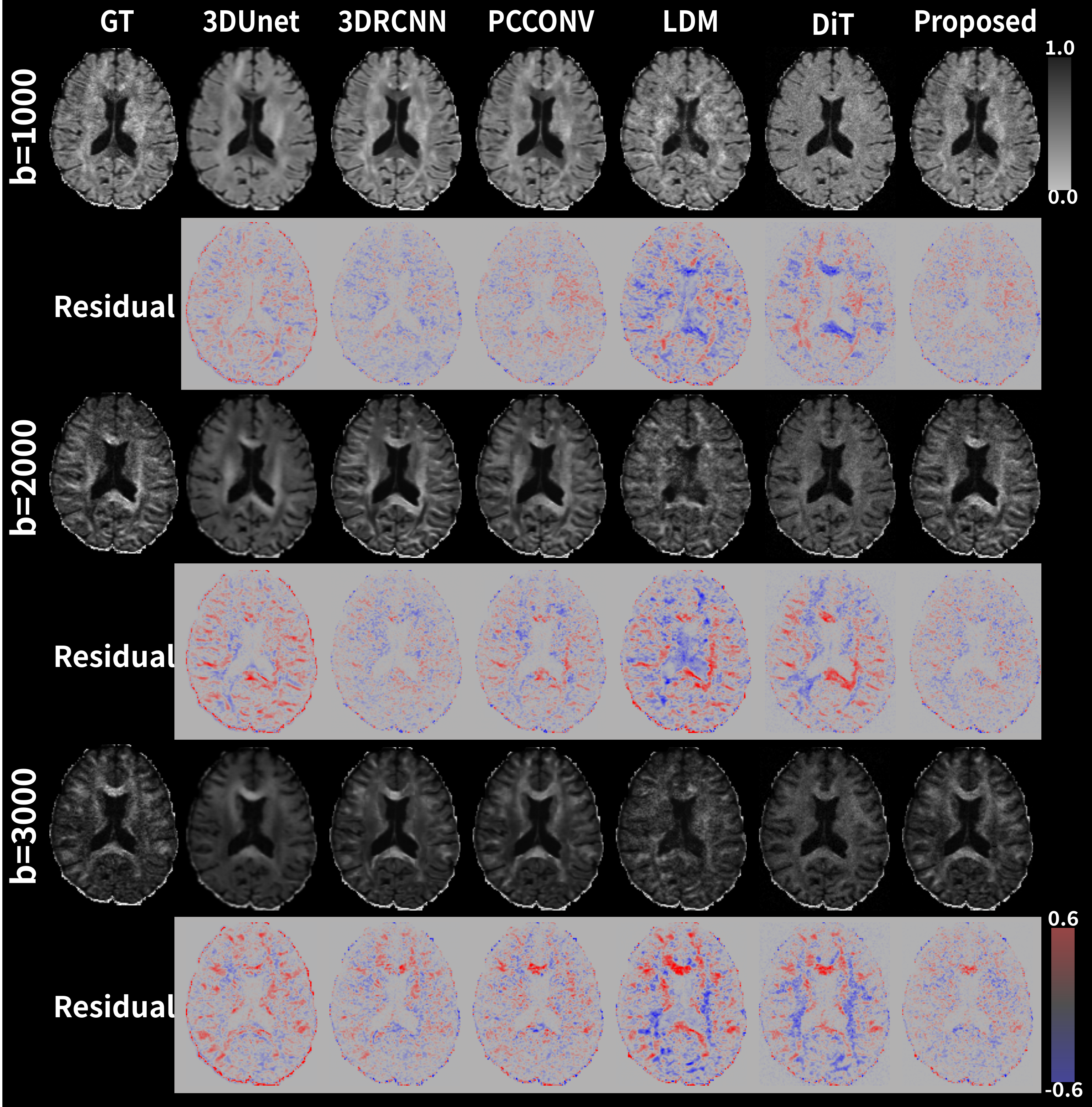}
\caption{The visualization results of HAR diffusion images obtained using different ASR methods with ASR scale $r=10$ and shells. Rows 2, 4, and 6 correspond to the respective error maps.}
\label{fig:dwi_brain}
\end{figure}

We evaluated the abovementioned representative general ASR methods and our proposed method on three ASR scales $r=15,9,6$ (6→90, 10→90, 15→90) across b=1000, 2000, and 3000 s/mm² shells. Table \ref{tab:dwi} shows the quantitative results of ASR diffusion images obtained from a random subject's middle slice by different methods. Vertically comparing, models that actively embed directional information, such as 3DRCNN, PCConv and our proposed method, demonstrate reduced residuals and enhanced reconstruction fidelity, while our method almost outperforms other methods in nearly every configuration, even at the $r = 15$, highlighting the effectiveness of our learning framework. Furthermore, from the voxel-wise scatter plots of Fig.\ref{fig:scatter}, our proposed method achieves the highest correlation, demonstrating superior reconstruction fidelity and angular agreement with the ground truth. From the qualitative results of Fig. \ref{fig:dwi_brain}, the visualization results and error maps demonstrate that our method outperforms others and restores more details. CNN-based methods, including 3DUNet, 3DRCNN, and PCConv, tend to produce overly smooth reconstructions, smoothing over high-frequency q-space variations. Diffusion-based models, LDM, DiT and our proposed method, exhibit stronger detail recovery, capturing noise-like textures and diffusion directional heterogeneity.To note, LDM performs worst on un-denoised HCP data, which might be  that the low-SNR DWI inputs results in degraded latent representations and inconsistent reconstructions. The original DiT, as our baseline model, also falls short in data fidelity without explicit directional conditioning or data consistency constraints. 

\begin{figure}[htbp]
\centering
\includegraphics[width=\linewidth, height=0.4\textheight, keepaspectratio]{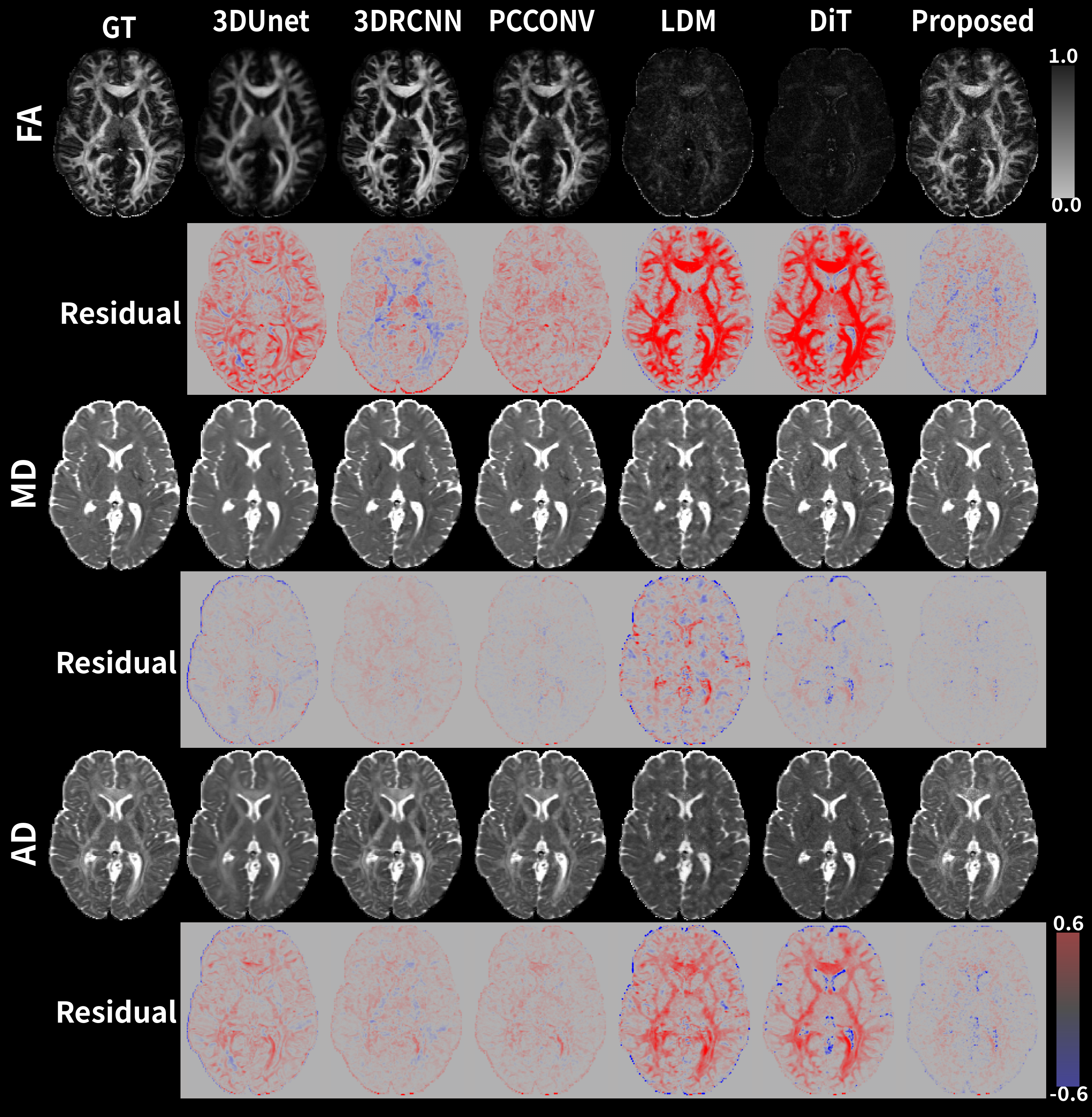}
\caption{Qualitative results of FA, MD and AD parameter maps obtained by different methods with ASR scale $r=15$. Rows 2, 4 and 6 present the corresponding error maps.}
\label{fig:dti_brain}
\end{figure}

\subsection{Downstream tasks}
\subsubsection{Comparison of DTI parameter maps}
\begin{figure*}[htbp]
\centering
\includegraphics[width=\linewidth, height=0.35\textheight,keepaspectratio]{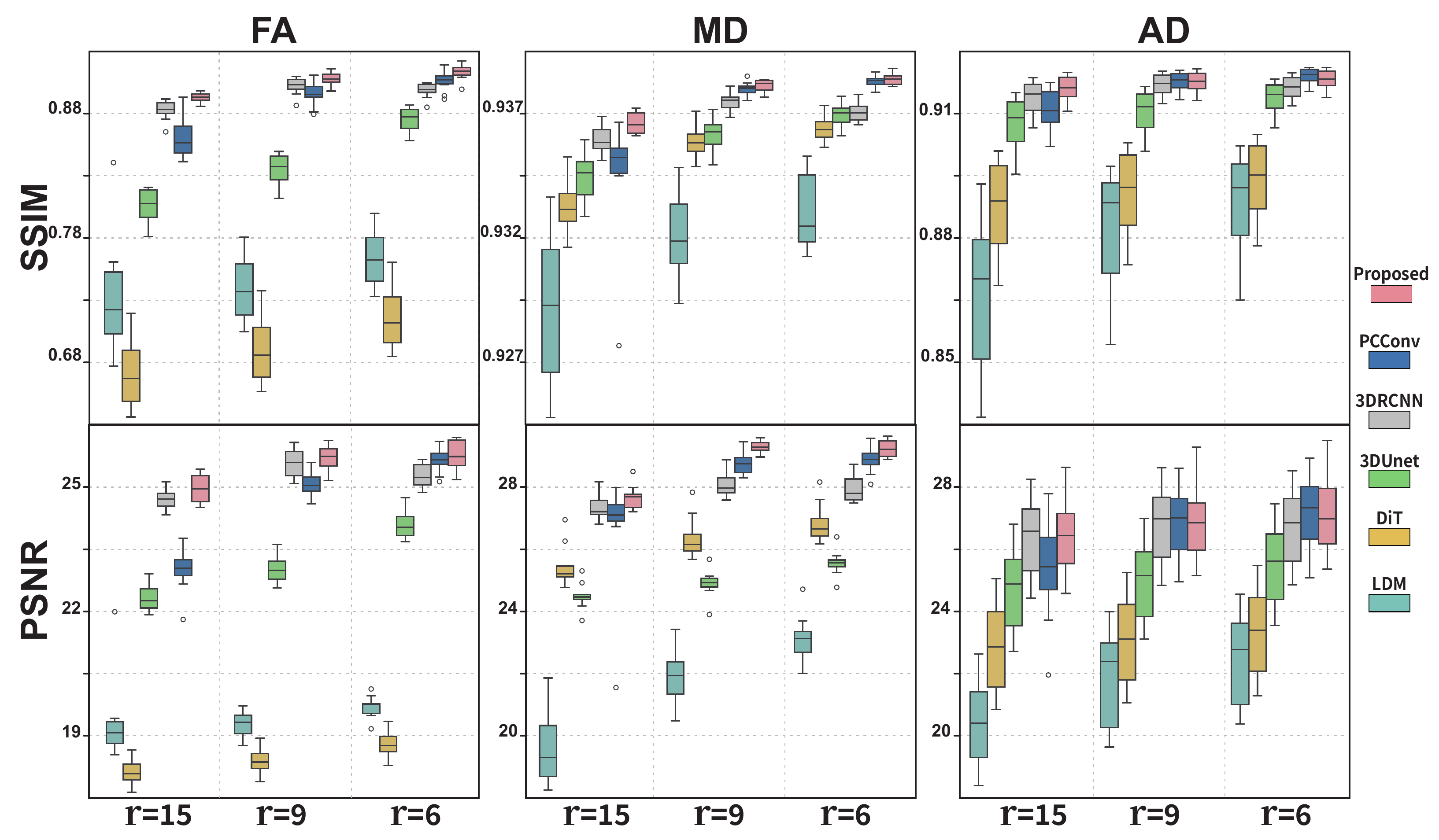}
\caption{
Boxplots of the distribution of SSIM and PSNR for reconstructed FA, MD, and AD across three ASR scales.
}
\label{fig:dti}
\end{figure*}

To further assess the clinical relevance of our model's performance, DTI parameters are used to quantify the proposed ASR method's ability to recover DTI‐derived microstructural metrics from undersampled data. All models are trained and evaluated on $b=1000s/mm^{2}$ data under ASR scale $r=15,9,6$. The reconstructed 90 direction data are used to fit a DTI model via DIPY\cite{garyfallidis2014dipy}, from which FA, MA and AD are computed. MA and AD are normalized to value range of 0,1 for consistent comparison. Fig \ref{fig:dti} presents the quantitative results of DTI-derived parameter maps. Across all scales and metrics, our proposed method consistently outperforms others. Notably, it achieves the strongest performance in FA reconstruction, which is known to be particularly sensitive to angular resolution and noise, highlighting our model's ability to preserve directional diffusivity information. Fig \ref{fig:dti_brain} visualizes the parameter maps generated by different methods under ASR scale $r=15$. The accompanying error maps demonstrate that our method yields superior reconstructions with improved detail preservation and noise suppression compared to other models.

Interestingly, all evaluated ASR methods exhibit a consistent performance hierarchy: MD is the easiest to recover, followed by AD, with FA being the most challenging. This trend aligns with prior findings showing that MD is inherently more robust to noise and requires less angular information for accurate estimation \cite{tae2018current}. AD, which captures diffusion along dominant axon bundles, is somewhat more sensitive but remains relatively robust due to its directional coherence. In contrast, FA, quantifying the shape asymmetry of the diffusion tensor, relies heavily on accurate angular modeling and is thus more vulnerable to sparse or noisy inputs \cite{seo2019reduction}. Consistent with this interpretation, even methods lacking explicit directional conditioning achieve relatively high SSIM scores for MD, while their FA estimates remain comparatively poor. Conversely, models incorporating directional embeddings—such as 3DRCNN, PCConv, and our proposed method—exhibit significant gains in FA reconstruction, underscoring the critical role of angular encoding mechanisms in improving microstructural fidelity.
\subsubsection{Comparison of NODDI parameter maps}
\begin{figure}[htbp]
\centering
\includegraphics[width=\linewidth, height=\textheight, keepaspectratio]{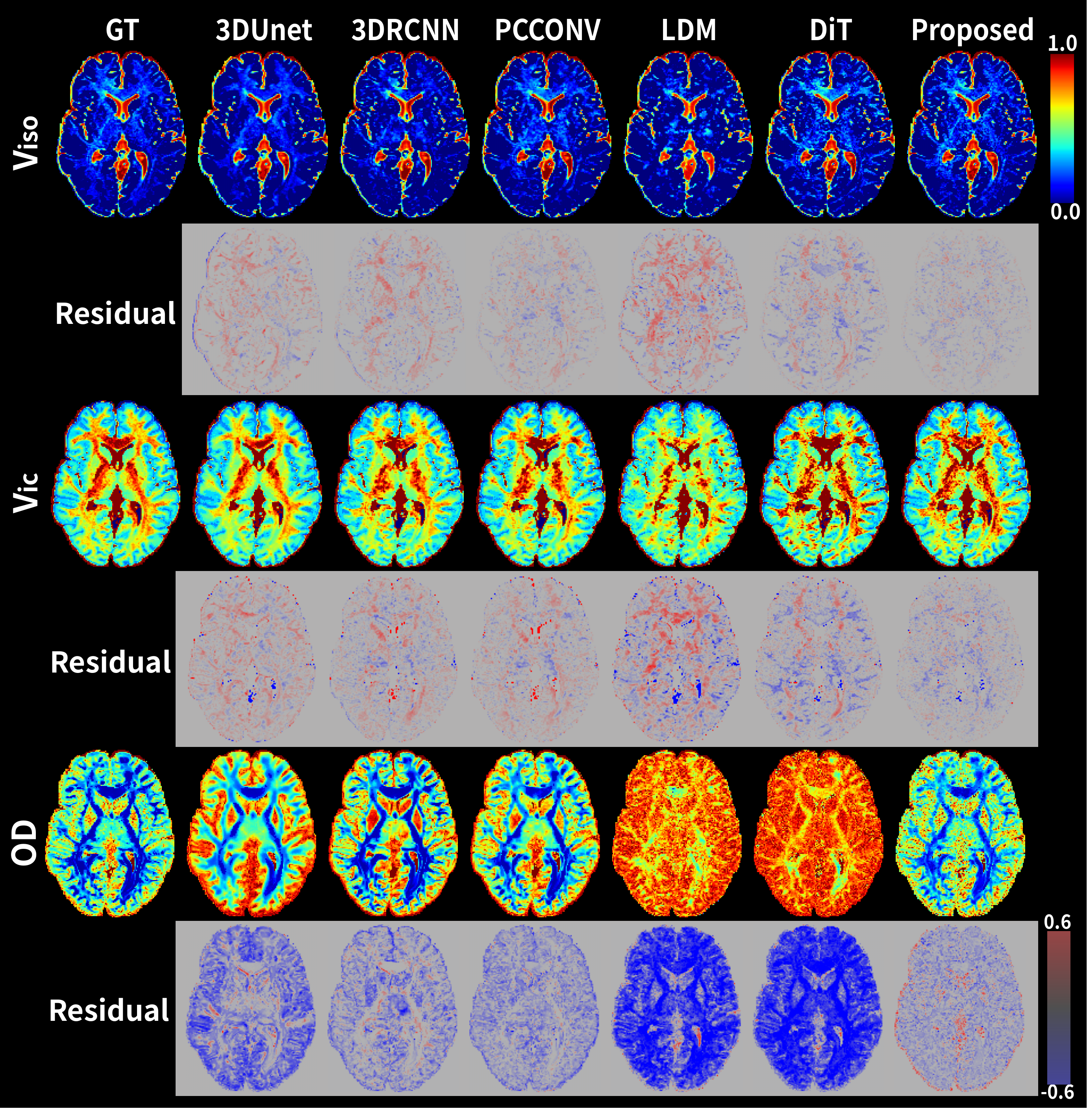}
\caption{Qualitative results of $V_{iso}$, $V_{ic}$ and OD parameter maps obtained by different methods with ASR scale $r=6$. Rows 2, 4 and 6 present the corresponding error maps.}
\label{fig:noddi_brain}
\end{figure}

\begin{figure*}[htbp]
\centering
\includegraphics[width=\linewidth, height=0.35\textheight,keepaspectratio]{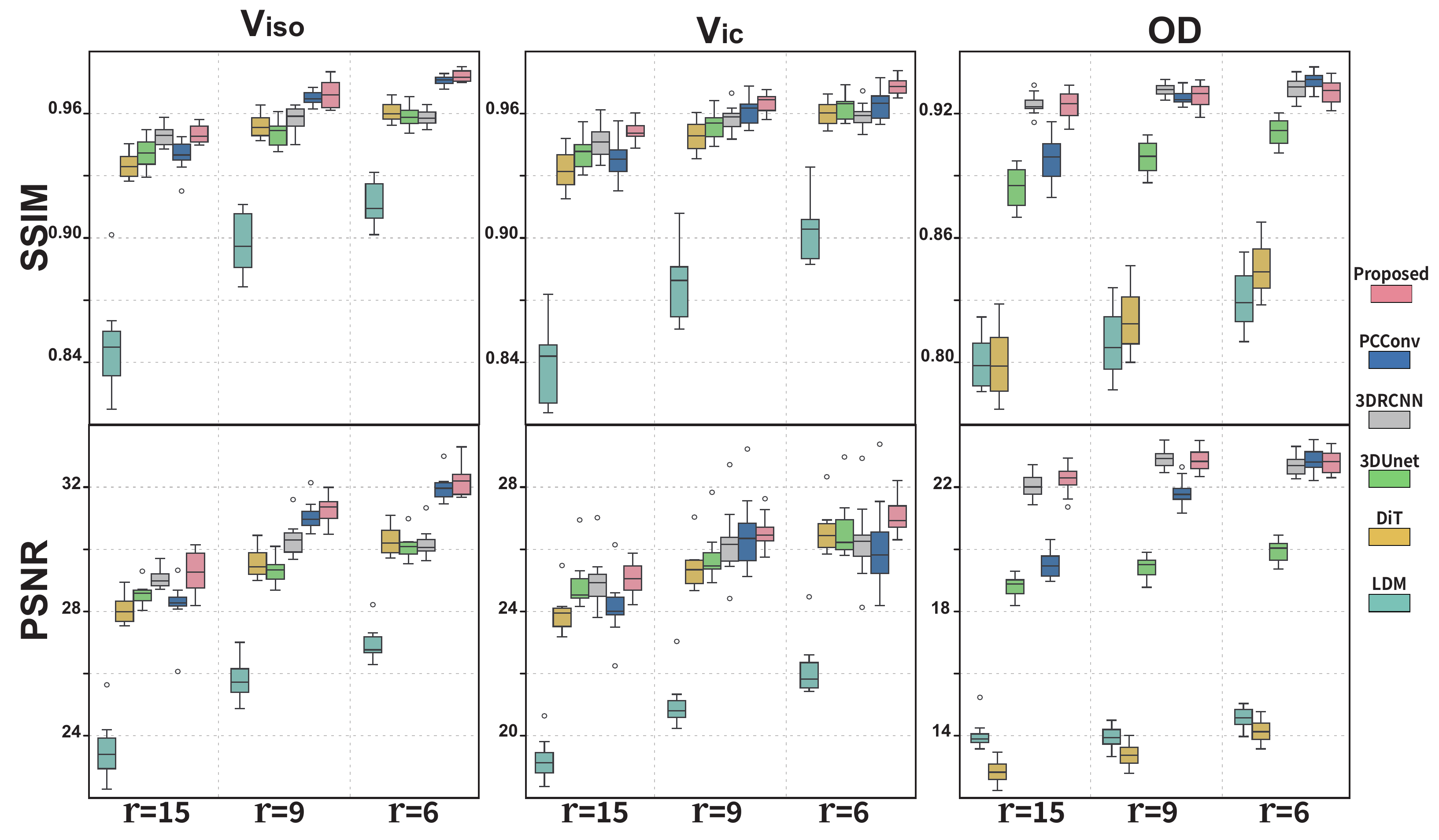}
\caption{
Boxplots of the distribution of SSIM and PSNR for reconstructed $V_{iso}$, $V_{ic}$ and OD across three ASR scale.
}
\label{fig:noddi}
\end{figure*}

We further evaluate the ASR methods in complex microstructural modeling. Specifically, for each model, we perform ASR scale of $r=15,10,6$ reconstruction independently on $b=1000s/mm^{2},2000,3000$ shells, respectively. For each ASR scale, the resulting three-shell super resolved $90×3=270$ reconstructed DWIs are then used to estimate NODDI microstructural parameters OD,$V_{ic}$ and $V_{iso}$ via AMICO\cite{daducci2015accelerated}. Quantitative results are shown in Fig \ref{fig:noddi}), and our method almost achieves the highest SSIM and PSNR while for all three NODDI metrics across every ASR ratio. Qualitative comparisons results of NODDI under ASR scale $r=6$ are displayed in Fig. \ref{fig:noddi_brain} supports these results.The reconstructed NDI and ODI maps produced by our method closely resemble the ground truth, with sharper delineation of cortical structures and finer white matter variation. The advantages are even more evident in the error maps: competing methods introduce substantial estimation errors and fail to capture anatomical detail. Several baseline models exhibit prominent artifacts or blurred boundaries around gray matter regions—likely caused by over-smoothing tendencies in CNN-based architectures. A similar performance hierarchy is observed in the NODDI metrics,  the $V_{iso}$ is the easiest to predict accurately, followed by $V_{ic}$ and then the OD being the most challenging. This trend reflects the increasing dependence of each parameter on directional detail and robustness to noise. $V_{iso}$ representing isotropic diffusion such as CSF, is largely orientation-invariant, $V_{ic}$ measures neurite density along dominant directions, while OD captures angular dispersion of neurites, making it highly sensitive to both angular resolution and fine structural features. These results demonstrate that our method not only excels in raw DWI ASR but also enhances the fidelity of higher-order biophysical estimations. This underscores its robustness across shells and its clinical viability for advanced diffusion modeling.

\subsection{Ablation study}
To rigorously quantify the contribution of each architectural enhancement, we compare four model variants: the original DiT baseline, DiT augmented with the Q-space Geometry-Aware Module (denoted as w/ QGAM), DiT with Spherical Harmonics–Guided Posterior Sampling (denoted as w/ SHPS), and our final proposed method. The ablation results, summarized in Table~\ref{tab:abla}, show consistent and interpretable improvements across all ASR scales and b-value shells as physics-informed components are progressively incorporated during both training and inference.
\begin{table*}[!ht]
\centering
\caption{THE ALATION STUDIES OF THE QGAM AND SHPS WERE CONDUCTED ACROSS THREE SHELLS and THREE ASR SCALES. BOLD NUMBERS INDICATE THE BEST RESULTS.}
\label{tab:abla}
\resizebox{0.95\textwidth}{!}{
\begin{tabular}{ccllllll}
\toprule[1pt]
\noalign{\vskip 0.5ex}
\textbf{b Value}&  & \multicolumn{2}{c}{$\mathbf{1000s/mm^{2}}$} & \multicolumn{2}{c}{$\mathbf{2000s/mm^{2}}$} & \multicolumn{2}{c}{$\mathbf{3000s/mm^{2}}$} \\[0.5ex] \hline
\noalign{\vskip 0.5ex}
\textbf{ASR Scale} & \textbf{Methods} & \multicolumn{1}{c}{\textbf{SSIM}} & \multicolumn{1}{c}{\textbf{PSNR}} & \multicolumn{1}{c}{\textbf{SSIM}} & \multicolumn{1}{c}{\textbf{PSNR}} & \multicolumn{1}{c}{\textbf{SSIM}} & \multicolumn{1}{c}{\textbf{PSNR}} \\[0.5ex] \midrule[1pt]
\noalign{\vskip 0.5ex}
 & Baseline & $0.9195\pm0.0060$ & $26.8805\pm0.3753$ & $0.8989\pm0.0101$ & $26.6302\pm0.4451$ & $0.8948\pm0.0099$ & $26.6136\pm0.5544$ \\
$r=15$ & w/ QGAM & $0.9534\pm0.0052$ & $28.3505\pm0.4010$ & $0.9478\pm0.0073$ & $27.9841\pm0.5076$ & $0.9317\pm0.0086$ & $27.5389\pm0.7212$ \\
 & w/ SHPS & $0.9374\pm0.0079$ & $27.7576\pm0.3437$ & $0.9232\pm0.0085$ & $27.2481\pm0.4930$ & $0.9165\pm0.0106$ & $26.9640\pm0.0432$ \\
 & Proposed & $\mathbf{0.9569\pm0.0054}$ & $\mathbf{28.5968\pm0.4655}$ & $\mathbf{0.9467\pm0.0065}$ & $\mathbf{28.0721\pm0.5739}$ & $\mathbf{0.9332\pm0.0081}$ & $\mathbf{27.4751\pm0.7392}$ \\[0.5ex] \midrule[1pt]
\noalign{\vskip 0.5ex}
 & Baseline & $0.9263\pm0.0056$ & $27.3481\pm0.3810$ & $0.9044\pm0.0094$ & $26.8650\pm0.4569$ & $0.9009\pm0.0092$ & $26.7294\pm0.5635$ \\
$r=9$ & w/ QGAM & $0.9585\pm0.0047$ & $28.8572\pm0.4420$ & $0.9493\pm0.0070$ & $28.1851\pm0.5417$ & $0.9355\pm0.0079$ & $27.5220\pm0.7187$ \\
 & w/ SHPS & $0.9436\pm0.0071$ & $28.0337\pm0.2961$ & $0.9293\pm0.0107$ & $27.5005\pm0.4876$ & $0.9196\pm0.0087$ & $27.0860\pm0.0430$ \\
 & Proposed & $\mathbf{0.9624\pm0.0049}$ & $\mathbf{29.0646\pm0.4832}$ & $\mathbf{0.9501\pm0.0071}$ & $\mathbf{28.3290\pm0.5462}$ & $\mathbf{0.9369\pm0.0080}$ & $\mathbf{27.5745\pm0.7126}$ \\[0.5ex] \midrule[1pt]
 \noalign{\vskip 0.5ex}
 & Baseline & $0.9311\pm0.0055$ & $27.6806\pm0.3787$ & $0.9117\pm0.0087$ & $27.0691\pm0.4654$ & $0.9071\pm0.0086$ & $26.9198\pm0.5789$ \\
$r=6$ & w/ QGAM & $0.9618\pm0.0053$ & $29.1446\pm0.3956$ & $0.9503\pm0.0068$ & $28.2475\pm0.5761$ & $0.9361\pm0.0082$ & $27.5604\pm0.7019$ \\
 & w/ SHPS & $0.9480\pm0.0047$ & $28.3405\pm0.0355$ & $0.9307\pm0.0084$ & $27.6227\pm0.0484$ & $0.9226\pm0.0076$ & $27.2187\pm0.0424$ \\
 & Proposed & $\mathbf{0.9655\pm0.0046}$ & $\mathbf{29.4183\pm0.4996}$ & $\mathbf{0.9520\pm0.0063}$ & $\mathbf{28.4335\pm0.6105}$ & $\mathbf{0.9386\pm0.0074}$ & $\mathbf{27.6289\pm0.6945}$ \\[0.5ex] \bottomrule[1pt]
\end{tabular}
}
\end{table*}
Compared to the baseline DiT model~\cite{peebles2023scalable}, the inclusion of QGAM substantially improves SSIM and PSNR across all conditions. In several cases, SSIM improves by up to 0.04, and PSNR increases by more than 2.0 dB. These improvements are consistent with prior findings that explicit conditioning on directional inputs enables the network to better align its internal representations with q-space geometry~\cite{perez2018film, liu2025frame}. Given that DWI signals vary significantly across directions due to underlying microstructural anisotropy~\cite{tournier2011diffusion}, direction-aware modulation helps disentangle angular dependencies and enhances sensitivity to orientation-specific features. As previously discussed, the DiT baseline lacks mechanisms for modeling q-space correlation and tends to ignore directional specificity. Consequently, it often performs better on isotropic parameters such as MD and $V_{iso}$, which reflect global diffusivity or free water content, but struggles with anisotropic metrics like FA and OD that require fine-grained modeling of directional diffusion behavior. The addition of SHPS further enhances robustness, particularly under low-SNR conditions. By introducing SH domain high-frequency regularization, SHPS provide a physically grounded inductive bias that guides DiT’s predictions toward the plausible data manifold, promoting smoother angular reconstructions and better continuity. While SHPS alone may underperform QGAM in some metrics, potentially due to its focus on global smoothness rather than localized directional adaptation, it proves especially effective at higher b-values, where noise dominates and statistical regularization becomes essential. Taken together, the results suggest that QGAM and SHPS are complementary: QGAM offers localized, direction-specific conditioning, while SHPS enforces global structural plausibility during inference. Their synergistic integration yields significant performance gains, particularly under sparse input and high b-value conditions (e.g.,$r=15$, $b = 3000$ s/mm$^2$),which are especially vulnerable to angular aliasing and signal degradation.

\section{Discussion}
This work introduces PGDiT, a physics-guided diffusion transformer framework that integrates q-space geometry–aware pretraining with spherical harmonics–guided sampling. Through extensive experiments on HCP data, we demonstrate its superiority in angular super-resolution, both in raw high angular-resolution dMRI reconstruction and in downstream microstructural analyses. The framework's principled design enables better performance across b-values and resilience to sparse sampling, making it a promising tool for efficient diffusion MRI acquisition and modeling. Future extensions could explore its future clinical application.

\bibliographystyle{IEEEtran}
\bibliography{PGDiT_TMI}

\end{document}